\documentclass[12pt]{iopart}

\usepackage{graphicx}
\newcommand{\ket}[1]{|{#1}\rangle}
\newcommand{\bra}[1]{\langle{#1}|}

\begin{document}

\title[Experimental ancilla-assisted qubit transmission]{Experimental ancilla-assisted qubit transmission against correlated noise using quantum parity checking}

\author{T Yamamoto$^{1,2}$, R Nagase$^{1,2}$, J Shimamura$^{1,2,3}$, \\
\c{S} K \"Ozdemir$^{1,2,3}$, M Koashi$^{1,2,3}$ and N Imoto$^{1,2,3}$}

\address{$^1$Department of Materials Engineering Science, Graduate school of Engineering Science, Osaka University, Toyonaka, Osaka 560-8531, Japan}
\address{$^2$CREST Research Team for Photonic Quantum Information, 4-1-8 Honmachi, Kawaguchi, Saitama 331-0012, Japan}
\address{$^3$SORST Research Team for Interacting Carrier Electronics, 4-1-8 Honmachi, Kawaguchi, Saitama 331-0012, Japan}
\ead{yamamoto@mp.es.osaka-u.ac.jp}
\begin{abstract}
We report the experimental demonstration of a transmission scheme of photonic qubits over unstabilized optical fibers, which has the ability to transmit any state of a qubit, regardless of whether it is known, unknown, or entangled to other systems. 
A high fidelity to the noiseless quantum channel was achieved by adding an ancilla photon after the signal photon within the correlation time of the fiber noise and by performing a measurement which computes the parity. Simplicity, maintenance-free feature and robustness against path-length mismatches among the nodes make our scheme suitable for multi-user quantum communication networks.

\end{abstract}

\maketitle

\section{Introduction}

Quantum communication networks with many participants will provide various communication and computation tasks based on the nature of quantum physics, such as quantum key distribution\cite{Bennett84,Ekert91PRL,Gisin02}, quantum teleportation\cite{Bennett93PRL},  quantum repeaters\cite{Briegel98}, measurement-based quantum computing\cite{Raussendorf01PRL}, 
and others\cite{Bennett92PRLa,Karlsson98PRA,Hillery99PRA}. 
Such a system inevitably involves manipulation of multi-partite entanglement, and requires faithful node-to-node transmission of an information carrier that is already entangled to other systems. Widespread use of such networks also demands a plug-and-play connectivity, which avoids the need for complicated stabilization and calibration tasks among distantly located users. 
Recent studies on practical quantum communication systems have mainly been focused on quantum key distribution (QKD).  Among the most promising implementations for QKD are the plug-and-play schemes\cite{Muller97,Stucki,Kwiat00S,Walton03PRL,Boileau04PRLb,Jiang05OE,Zhang06PRA,Chen06PRL} 
and those utilizing double Mach-Zender interferometers (MZI)\cite{Bennett92PRL,Gobby04,Kimura04},  
both of which share the common feature of robustness against correlated noise during transmission of quantum states in the optical fibers. In the plug-and-play systems based on the auto-compensation of birefringence effects during a round trip of light pulses\cite{Muller97,Stucki}, 
the encoding of the transmitted states is done by choosing a manipulation on the incoming pulse, which implies that the transmitted state must be known to the sender. The plug-and-play feature can be also achieved by utilizing multi-photon entangled states in decoherence-free subspaces(DFS)\cite{Kwiat00S,Walton03PRL,Boileau04PRLb,Jiang05OE,Zhang06PRA,Chen06PRL}. 
In this case, transmission of a photonic qubit in an unknown state requires an encoding process into multi-photon entangled states, which is difficult in the present technology. On the other hand, double MZI systems can be used for arbitrary quantum states whether it is known or unknown; however, the need for subwavelength-optical-delay adjustments in MZIs at each node stands as a disadvantage, especially when the number of the participants in the network increases. The existing schemes therefore lack either the plug-and-play feature or the ability to transmit a qubit that is in an unknown state or is entangled to other systems. Achieving both of the features at the same time is not only of practical importance but also of fundamental interest, since it amounts to a faithful transmission of quantum states between the parties who do not have the shared reference frame\cite{Bartlett03PRL}. 

\begin{figure}[tbp]
 \centering
 \includegraphics[scale=1]{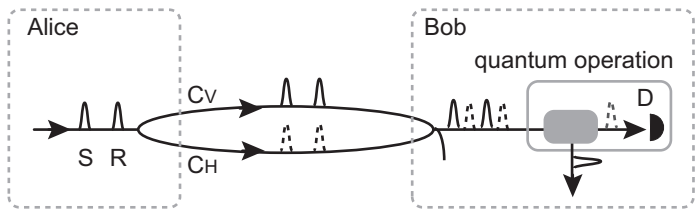}
 \caption {
Concept of ancilla-assisted faithful transmission of photonic qubit state.
Signal (S) and reference (R) photons are sequentially transmitted from Alice to Bob through dual-rail quantum channel composed of two optical fibers C$_{\rm H}$ and C$_{\rm V}$, each adding identical fluctuations to the photons. Bob extracts the signal photon from the received two photons by a quantum operation which can be implemented by linear optical elements and a photon detector D as shown in fig.~2. 
} \label{fig:setup}
\end{figure}

 In this letter, we experimentally demonstrate such a faithful transmission scheme fulfilling  both of the above requirements, for single-photon polarization states through optical fibers. 
This is achieved by adding an ancilla photon of a fixed polarization after the signal photon within the correlation time of the phase fluctuations in the fiber and by quantum parity checking. After the transmission of these photons through unstabilized optical fibres, the channel fidelity to the noiseless quantum channel is $0.958$. Evidently, a transmission channel which is very close to a noiseless one is achieved without the stabilization of optical components.

\section{Ancilla-assisted qubit transmission scheme}
We first introduce the idea\cite{TYamamoto05PRL} of the scheme and then describe the experimental demonstration using two photons generated by spontaneous parametric down-conversion (SPDC), the linear optical elements and the photon detectors.

Suppose that Alice is given a  signal photon in unknown state $\alpha\ket{\rm H}+\beta\ket{\rm V}$, where $\ket{\rm H}$ and $\ket{\rm V}$ represent horizontal (H) and vertical (V) polarization states, respectively, and $|\alpha|^2+|\beta|^2=1$. Alice uses another photon as a reference, which she prepares in a fixed state $\ket{\rm D}\equiv (\ket{\rm H}+\ket{\rm V})/\sqrt{2}$.
She sends the signal photon in a time-bin following that of the reference photon with a temporal delay $\Delta t_{\rm A}$.
The two-photon state can be written as
\begin{eqnarray}
\ket{\rm D}\otimes(\alpha\ket{\rm H}_{\Delta t_{\rm A}}+\beta\ket{{\rm V}}_{\Delta t_{\rm A}}),
\label{eq:1}
\end{eqnarray}
where the subscripts represent the temporal delay from the front time-bin.
As shown in Fig.~1, the photons in the H- and the V-polarization state are transmitted through the channels C$_{\rm H}$ and C$_{\rm V}$, respectively. While ordinary single-mode fibers can be used for these quantum channels at the cost of decreasing the success probability\cite{TYamamoto05PRL}, 
here we use polarization-maintaining optical fibers (PMF) for the simplicity of the experiments.
In this case, the polarization rotations of the photons in each channel do not occur, but unknown phase shifts $\phi_{\rm H}$ and $\phi_{\rm V}$ are added to the photons in
each channel independently due to the fluctuations of the optical path lengths.
We assume the interval $\Delta t_{\rm A}$ between the signal and reference photons is much shorter than the correlation time of the fluctuations, so that the phase shifts are considered to be correlated such as $\phi_{\rm H}(t)=\phi_{\rm H}(t+\Delta t_{\rm A})=\phi_{\rm H}$ and $\phi_{\rm V}(t)=\phi_{\rm V}(t+\Delta t_{\rm A})=\phi_{\rm V}$.
At Bob's location, the photons in both modes, C$_{\rm H}$ and C$_{\rm V}$, are mixed together, and
 the received state becomes
\begin{eqnarray}
& &1/\sqrt{2}[\alpha e^{2i\phi_{\rm H}}\ket{\rm H}\ket{\rm H}_{\Delta t_{\rm A}}+\beta e^{2i\phi_{\rm V}}\ket{{\rm V}}_{\tau}\ket{{\rm V}}_{\Delta t_{\rm A}+\tau} \nonumber \\
& &+e^{i(\phi_{\rm H}+\phi_{\rm V})}(\alpha \ket{{\rm V}}_{\tau}\ket{{\rm H}}_{\Delta t_{\rm A}}+\beta \ket{{\rm H}}\ket{{\rm V}}_{\Delta t_{\rm A}+\tau})].
\label{eq:3}
\end{eqnarray}
Here the optical path lengths of C$_{\rm V}$ and C$_{\rm H}$ may differ, which is indicated by the temporal delay $\tau$ in the subscripts of the V-polarization states.
We can easily see that the state $\alpha \ket{{\rm V}}_{\tau}\ket{{\rm H}}_{\Delta t_{\rm A}}+\beta \ket{{\rm H}}\ket{{\rm V}}_{\Delta t_{\rm A}+\tau}$ is invariant under the phase shifts. 
 It is worth mentioning that in the previous DFS schemes \cite{Kwiat00S,Walton03PRL,Boileau04PRLb,Jiang05OE,Zhang06PRA,Chen06PRL}, Alice prepares entangled states in the DFS. In our scheme, Alice's two photons [Eq. (1)] are not correlated, let alone entangled. It is Bob who sifts out the entangled states in the DFS. Bob can, in principle, project the state (\ref{eq:3}) onto  the state $\alpha \ket{{\rm V}}_{\tau}\ket{{\rm H}}_{\Delta t_{\rm A}}+\beta \ket{{\rm H}}\ket{{\rm V}}_{\Delta t_{\rm A}+\tau}$, which happens with the probability of $1/2$, and decode the projected state into the faithful signal $\alpha \ket{{\rm H}}+\beta \ket{{\rm V}}$. In our experiment, this extraction of the faithful signal state from the received state is performed by passive linear optical elements and postselection using the photon detectors. 

\begin{figure}[tbp]
 \centering
\includegraphics[scale=1]{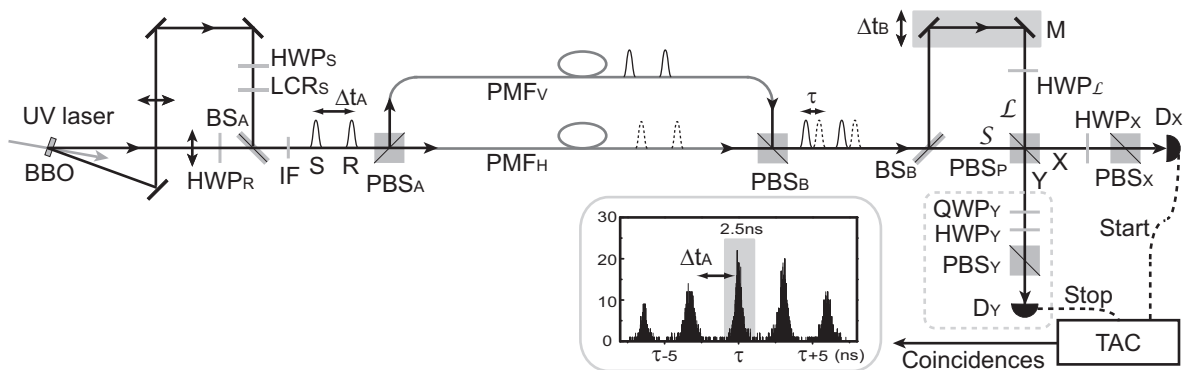}
 \caption {
Experimental set-up composed of two-photon generation, linear optics and photon detectors.
The UV light beam (average power 280 mW) used for pumping the BBO crystal for SPDC is obtained from a frequency doubled mode-locked Ti:sapphire laser (wavelength: 790 nm, pulse width: 80 fs, repetition rate: 82 MHz). The spectral filtering of the generated photons is performed by narrow band interference filter (IF, wavelength: 790 nm, band width: 3.5 nm). 
All detectors D$_{\rm X}$ and D$_{\rm Y}$ are silicon avalanche photodiodes, and they are placed after single-mode optical fibers. 
The histogram shows the number of delayed coincidence events with various delay between the detectors D$_{\rm X}$ and D$_{\rm Y}$, which was recorded by time-to-amplitude converter (TAC). 
The central peak shows the events where the signal and reference photons have passed $\cal S$ and $\cal L$, respectively. We accept the events in 2.5 ns time window around the central peak as the successful ones. 
Note that the two peaks separated from the central peak by $\Delta t_A$ correspond to the case where both photons pass through $\cal S$ or $\cal L$, and the remaining two other peaks correspond to the case where the signal passes $\cal L$ and the reference passes $\cal S$.
 } \label{fig:setupEx}
\end{figure}

\section{Experiment}
The schematics of the experimental set-up is shown in Fig.~2. Two photons in distinct modes are generated by SPDC from Type I phase matched 2-mm-thick $\beta$-barium borate (BBO) crystals. One photon in $\ket{{\rm H}}$ passes through long path, and is transformed into arbitrary signal polarization states $\alpha\ket{{\rm H}}_{\Delta t_{\rm A}}+\beta\ket{{\rm V}}_{\Delta t_{\rm A}}$ by rotating the polarization by a half wave plate HWP$_{\rm S}$ and adding a phase shift by a liquid crystal retarder LCR$_{\rm S}$.
The other photon in $\ket{{\rm H}}$ passes through short path, and is transformed into the fixed reference polarization state by  HWP$_{\rm R}$. 
These photons are mixed by a non-polarizing beamsplitter BS$_{\rm A}$. Here we can prepare the two photons in the state~(\ref{eq:1}) with the probability 1/4 when two photons are generated from SPDC. The temporal delay $\Delta t_{\rm A}$ between the signal and the reference photon is about $3~{\rm ns}$. The photons are split into the H- and V-polarization modes by a polarizing beam splitter PBS$_{\rm A}$ which transmits the H-polarization photons and reflects the V-polarization photons. These photons are then transmitted to Bob through 10-m PMF$_{\rm H}$ and PMF$_{\rm V}$.

\begin{figure}[tb]
 \centering
 \includegraphics[scale=1]{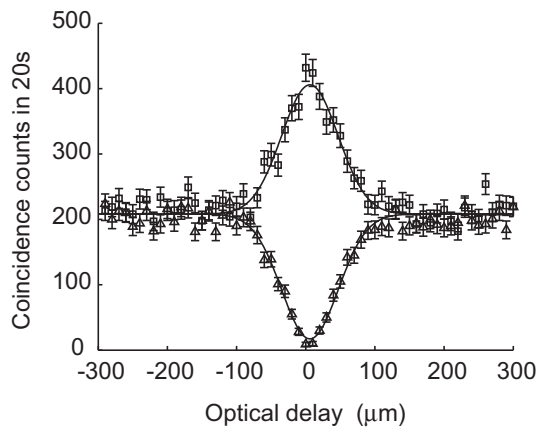}
 \caption {
\noindent 
Observed two photon interference as a function of optical delay $\Delta t_{\rm B}$.
The two markers show the coincidence counts measured on the bases $\ket{\bar{\rm D}}^{\rm Y}$(triangle) and $\ket{\rm D}^{\rm Y}$ (square) when the signal state is $\ket{\rm D}_{\Delta t_{\rm A}}$.
The error bars assume the Poisson statistics of the events. 
The solid Gaussian curves, which show the coherence length $l_c$(FWHM) $\simeq$ 75 $\rm \mu m$, represent the best fit to the data.
} \label{fig:lo}
\end{figure}

At Bob's location, these photons are mixed by PBS$_{\rm B}$ again. If the optical path lengths of PMF$_{\rm H}$ and PMF$_{\rm V}$ were precisely adjusted with high stability, the received state would be the same as the state prepared by Alice. However, we did not perform any such stabilizations in the following experiments taking several hours of data accumulation, during which the phase shifts $\phi_{\rm H}$ and $\phi_{\rm V}$ fluctuated randomly.

The extraction of the signal state from the received two-photon state can be passively performed in the following way. The received two photons are first split into long path ($\cal L$) and short path ($\cal S$) by BS$_{\rm B}$, then mixed by PBS$_{\rm P}$ again. HWP$_{\cal L}$ rotates the polarization of the photons in the long path by 90$^\circ$.  Using HWP$_{\rm X}$,  PBS$_{\rm X}$, and a photon detector D$_{\rm X}$, the polarization of the photon in mode X is projected onto the diagonal state $\ket{\rm D}$. The difference between the lengths of $\cal L$ and  $\cal S$ corresponds to a temporal delay $\Delta t_{\rm B}$ which is adjusted by the mirrors (M) on a motorized stage. The successful events are postselected by discriminating the time delay between the arrival of photons at detectors D$_{\rm X}$ and D$_{\rm Y}$ by using the time resolving coincidence detection as shown in Fig~2.  HWP$_{\rm Y}$ and the quarter wave plate QWP$_{\rm Y}$ in front of D$_{\rm Y}$ are used for the analyses of the successfully extracted signal states.

Here we only consider the successful case where the signal photon passes through $\cal S$ and the reference photon passes through $\cal L$. This happens with the probability 1/4 when two photons  arrived at the BS$_{\rm B}$.
In this case the state just before the PBS$_{\rm P}$ can be written as
\begin{eqnarray}
\alpha e^{2i\phi_{\rm H}}\ket{{\rm V}}^{\cal L}_{\Delta t_{\rm B}}\ket{{\rm H}}^{\cal S}_{\Delta t_{\rm A}}+\beta e^{2i\phi_{\rm V}}\ket{{\rm H}}^{\cal L}_{\tau+\Delta t_{\rm B}}\ket{{\rm V}}^{\cal S}_{\Delta t_{\rm A}+\tau} \nonumber \\
+e^{i(\phi_{\rm H}+\phi_{\rm V})}(\alpha \ket{{\rm H}}^{\cal L}_{\tau+\Delta t_{\rm B}}\ket{{\rm H}}^{\cal S}_{\Delta t_{\rm A}}+\beta \ket{{\rm V}}^{\cal L}_{\Delta t_{\rm B}}\ket{{\rm V}}^{\cal S}_{\Delta t_{\rm A}+\tau}),
\label{eq:5}
\end{eqnarray}
where the superscripts represent the spatial modes.
Here $\phi_{\rm H}$ and $\phi_{\rm V}$ include the phase shifts added in Bob's interferometer. If one photon is emitted in each mode of X and Y,  the output state just after the PBS$_{\rm P}$ is $\alpha \ket{{\rm H}}^{\rm X}_{\Delta t_{\rm A}}\ket{{\rm H}}^{\rm Y}_{\tau +\Delta t_{\rm B}}+\beta \ket{{\rm V}}^{\rm X}_{\Delta t_{\rm B}}\ket{{\rm V}}^{\rm Y}_{\Delta t_{\rm A}+\tau}$. This operation is referred to as quantum parity checking \cite{Pittman01PRA},  
which is also useful for other quantum information tasks \cite{TYamamoto03,Pan03}. 
Let us consider the case where $\Delta t_{\rm A}=\Delta t_{\rm B}$.
When the detector D$_{\rm X}$ finds one photon, the state in mode X is projected onto $\ket{\rm D}^{\rm X}_{\Delta t_{\rm A}}$ . At that time, the state in mode Y is projected onto the state
$\alpha \ket{{\rm H}}^{\rm Y}_{\tau +\Delta t_{\rm A}}+\beta \ket{{\rm V}}^{\rm Y}_{\Delta t_{\rm A}+\tau}$,
implying that 
we faithfully obtain the signal state in mode Y. It is worth to mention here that the delay $\tau$ affects only the arrival time but not the fidelity 
of the output states as long as the correlation time of the fluctuations of the phase shifts added in Bob's interferometer is much longer than $\tau$.

We first show that the above scheme can extract a faithful signal state in mode Y by properly adjusting the optical delay $\Delta t_{\rm B}$, when the signal state is $\ket{\rm D}_{\Delta t_{\rm A}}$. As shown in Fig.~3, varying the optical delay by moving M, we can clearly see the interference effects. The upper and lower curves show the coincidence rates on the bases $\ket{\rm D}^{\rm Y}$ and $\ket{\bar{\rm D}}^{\rm Y}\equiv (\ket{\rm H}^{\rm Y}-\ket{\rm V}^{\rm Y})/\sqrt{2}$, respectively.  The observed visibility at the zero delay is $0.959\pm 0.013$ representing a clear signature that coherence is preserved during quantum state transmission. The small deviation from 100\%  
visibility is due to the residual mode mismatch as well as 
multi-photon-pair generation during the SPDC. The full-width at 
half-maximum (FWHM) of the interference fringe, which corresponds 
to the coherence length $l_{\rm c}$ of the photons, is found to be $\sim 75$ $\rm 
\mu m$. This is roughly 100 times larger than the wavelength of 
the photons implying the robustness of the scheme against 
path-length mismatches and fluctuations up to the order of many 
wavelengths. 
The requirement for the precision of alignment and stability will be further relaxed if we choose the photons with longer $l_{\rm c}$.

\begin{figure}[tbp]
 \centering
 \includegraphics[scale=1]{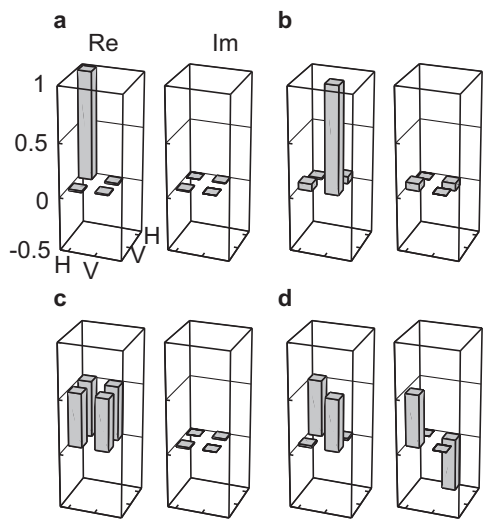}
 \caption { 
Experimental results showing that faithful qubit transmission is achieved.
 Real and imaginary components of density matrices  of the extracted states at Bob's side when the signal state is {\bf a} $\ket{\rm H}_{\Delta t_{\rm A}}$, {\bf b} $\ket{\rm V}_{\Delta t_{\rm A}}$, {\bf c} $\ket{\rm D}_{\Delta t_{\rm A}}$, and {\bf d} $\ket{\rm L}_{\Delta t_{\rm A}}$.
 Reconstruction is done by recording coincidence counts on four different settings of QWP$_{\rm Y}$ and HWP$_{\rm Y}$ in 100 s interval. 
 } \label{fig:fig3}
\end{figure}

In order to characterize the performance of our transmission scheme precisely, we analyzed the output states via tomographic reconstruction of the density matrices for various signal states.
Real and imaginary components of the density matrices of the output states are reconstructed for the input signal states, $\ket{ {\rm H}}_{\Delta t_{\rm A}}$, $\ket{ {\rm V}}_{\Delta t_{\rm A}}$, $\ket{\rm D}_{\Delta t_{\rm A}}$, and $\ket{\rm L}_{\Delta t_{\rm A}}\equiv (\ket{{\rm H}}_{\Delta t_{\rm A}}+i\ket{{\rm V}}_{\Delta t_{\rm A}})/\sqrt{2}$, and the results are shown in Fig.~4.
The fidelities of these reconstructed density matrices to those of the initial signal states are calculated as, $0.991\pm 0.031$, $0.985\pm 0.030$, $0.999\pm 0.030$, and $0.985\pm 0.030$, respectively, which clearly shows that the output states are very close to the input signal states.

Since the above experimental results are enough to characterize completely the quantum operation ${\cal E}$ effectively applied in our system, we can calculate various quantities for the demonstrated operation ${\cal E}$. In order to characterize quantitatively how close the operation ${\cal E}$ is to the noiseless quantum channel, we calculate the average fidelity $\bar{F}({\cal E})$ \cite{Schumacher96PRA} 
which is defined as the average of the fidelites $F_{i}=\bra{\psi_{i}}{\cal E}(\psi_{i})\ket{\psi_{i}}$ over all input states $\ket{\psi_{i}}$. It has been shown that $\bar{F}({\cal E})$ is connected to entanglement fidelity $F_{\rm e}({\cal E})\equiv \bra{\phi}({\cal I}_{\rm R}\otimes {\cal E})(\phi)\ket{\phi}$ by the following simple formula $\bar{F}({\cal E})=(2F_{\rm e}({\cal E})+1)/3$ for qubit channels \cite{Horodecki99PRA,Nielsen02PL}, 
where $\ket{\phi}$ represent a Bell state.  The ${\cal E}$ acts on one member of the Bell state $\ket{\phi}$ and ${\cal I}_{\rm R}$ acts on the other. Instead of measuring $F_{\rm e}({\cal E})$ by preparing the Bell state $\ket{\phi}$, we can estimate $F_{\rm e}({\cal E})$ from the above reconstructed density matrices as $0.958\pm 0.033$ and $\bar{F}({\cal E})$ is calculated to be $0.972\pm 0.022$. This clearly shows that our qubit transmission scheme provides a high fidelity to noiseless quantum channel.

The proof-of-principle experiment demonstrated here uses passive linear optical elements, thus the probability of success is rather small than the ideal case. However the success probability will increase 16 times by replacing the BS$_A$ and BS$_B$ by fast optical switches, and twice by using feed-forward decoding techniques. 
 Excluding the fiber losses and the detection loss, these improvements will enable  a success probability of 1/2. In the present experiment, we used PMFs as a channel,  but our scheme also allows the use of standard single-mode fibers\cite{TYamamoto05PRL}. 
It is worth noting that the photon losses in optical fibers may affect the efficiency in this two-photon quantum communication scheme more  than in the single photon transmission, but the efficiency will be greatly improved by the use of quantum repeaters\cite{Briegel98}.

\section{Conclusion}
This work demonstrates that two-photon interference effect together with quantum parity checking can be used for faithful transmission of qubits in arbitrary unknown quantum states with the help of ancillas without active control and stabilization mechanisms. Simplicity, versartaility, and maintenance-free feature of the scheme will be important for future quantum communication networks.

\section{Acknowledgements}
This work was supported by 21st Century COE Program by the Japan Society for the Promotion of Science and a MEXT Grant-in-Aid for Young Scientists (B) 17740265.

\end{document}